\pdfoutput=1 

\documentclass[11pt,a4paper]{article}

\usepackage{dcmi}

\usepackage[utf8]{inputenc} 
\usepackage[T1]{fontenc}    
\usepackage{hyperref}       
\usepackage{url}            
\usepackage{booktabs}       
\usepackage{amsfonts}       
\usepackage{nicefrac}       
\usepackage{microtype}      
\usepackage[pdftex]{graphicx}       
\usepackage{sectsty}		

\usepackage{natbib}

\usepackage{times}
\usepackage[utf8]{inputenc}
\usepackage{latexsym}
\usepackage{graphicx}
\usepackage{multirow}
\usepackage{listings}
\usepackage{listingsutf8}
\usepackage[table]{xcolor}
\usepackage{url}
\usepackage{hyperref}
\usepackage{tabularx}
\usepackage{threeparttable}

\usepackage{tikz}
\usetikzlibrary{calc}
\usepackage{algorithm}
\usepackage[noend]{algpseudocode}
\usepackage{amsmath}

\usepackage{subfigure}
\usepackage{svg}

\usepackage{xcolor}
\usepackage{pgfplots}
\usepackage[normalem]{ulem}

\newcommand{\grey}[1]{\textcolor{gray}{#1}}

\newcommand\capmystring[1]{\capmystringaux#1\relax}
\def\capmystringaux#1#2\relax{\uppercase{#1}\lowercase{#2}}
\newcommand{\xmltag}[1]{\emph{\capmystring{#1}}}
\newcommand{\xmlval}[1]{``\emph{#1}''}

\usepackage[strings]{underscore}
\usepackage[american]{babel}
\usepackage{csquotes}

\graphicspath{{Figures/}}

\newlength\myfntht
\newcommand{\percentbar}[3]{\setlength\myfntht{#2}
    #3~\makebox[\myfntht][l]{\rlap{\textcolor{black!20}{\rule[-1pt]{\myfntht}{1.9ex}}}\rule[-1pt]{#1\myfntht}{1.9ex}}}

\allsectionsfont{\sffamily}
\sectionfont{\sffamily\fontsize{12}{15}\selectfont}
\subsectionfont{\sffamily\fontsize{11}{14}\selectfont}
\subsubsectionfont{\sffamily\fontsize{10}{14}\selectfont\itshape}

\lstdefinelanguage{json}{
    basicstyle=\small\ttfamily,
    numberstyle=\scriptsize,
    stepnumber=1,
    numbersep=8pt,
    showstringspaces=false,
    breaklines=true,
    frame=lines,
    backgroundcolor=\color{background},
    moredelim=**[is][\color{forestG}]{@}{@},
    literate=
      {:}{{{\color{punct}{:}}}}{1}
      {,}{{{\color{punct}{,}}}}{1}
      {\{}{{{\color{delim}{\{}}}}{1}
      {\}}{{{\color{delim}{\}}}}}{1}
      {[}{{{\color{delim}{[}}}}{1}
      {]}{{{\color{delim}{]}}}}{1},
}

\usepackage{color}
\usepackage{inconsolata}
\definecolor{mauve}{HTML}{810381}

\lstdefinelanguage{CustomXML}
{
  morestring=[b]",
  moredelim=[s][\color{black}]{>}{<},
  moredelim=[s][\color{black}\textbf]{\ }{=},
  stringstyle=\color{blue},
  identifierstyle=\bfseries\color{mauve},
  basicstyle=\ttfamily\scriptsize\color{black}\selectfont
}

\definecolor{lightgray}{gray}{0.9}

\title{The state of OAI-PMH repositories\\in Canadian Universities}

\author{%
  Frédéric Piedboeuf\,\textsuperscript{\normalfont\dag}, Guillaume Le Berre\,\textsuperscript{\normalfont\dag}, David Alfonso-Hermelo\,\textsuperscript{\normalfont\dag},\\\textbf{Olivier Charbonneau\,\textsuperscript{\normalfont\ddag}, Philippe Langlais\,\textsuperscript{\normalfont\dag}} \\
  \textsuperscript{\normalfont\dag}\,DIRO, University of Montreal \\
  \textsuperscript{\normalfont\ddag}\,Concordia University \\
  Montreal, QC, Canada \\
  \texttt{\{frederic.piedboeuf, guillaume.le.berre, philippe.langlais\}@umontreal.ca}, \\ \texttt{davidalfonsohermelo@gmail.com, o.charbonneau@concordia.ca} \\
}

\begin{document}

\maketitle

\begin{abstract}
  This article presents a study of the current state of Universities Institutional Repositories (UIRs) in Canada. UIRs are vital to sharing information and documents, mainly Electronic Thesis and Dissertation (ETDs), and theoretically allow anyone, anywhere, to access the documents contained within the repository. Despite calls for consistent and shareable metadata in these repositories, our literature review shows inconsistencies in UIRs, including incorrect use of metadata fields and the omission of crucial information, rendering the systematic analysis of UIR complex. Nonetheless, we collected the data of 57 Canadian UIRs with the aim of analyzing Canadian data and to assess the quality of its UIRs. This was surprisingly difficult due to the lack of information about the UIRs, and we attempt to ease future collection efforts by organizing vital information which are difficult to find, starting from addresses of UIRs. We furthermore present and analyze the main characteristics of the UIRs we managed to collect, using this dataset to create recommendations for future practitioners.
\end{abstract}

\keywords{metadata; canada; institutional repositories; universities; OAI-PMH protocol; scholarly communication; digital scholarship; electronic thesis and dissertation; EDT}

\setcounter{footnote}{0}

\section{Introduction}

The term ETD, or \textit{Electronic Thesis or Dissertation}, refers to theses which are present in digital form, often online and in open access format. ETDs find themselves in document repositories of universities, with the associated metadata commonly entered by authors themselves, which saves considerable human and financial resources of cataloguing~\citep{robinson2016leveraging, reeves2007user}. Following open-access papers from journal and conferences, ETDs are estimated to be the most important document types in open archives~\citep{schopfel2013adding}. Furthermore, the existence of centralized repositories of research publications eases the analysis of research trends over time~\citep{gokmen2017methodological}, and has also allowed the emergence of \textit{bibliomining}, the study of documentalist tendencies using mined data from repositories~\citep{siguenza2015literature}. 

Canada's presence in the international research scene is a large one, which is illustrated by its coming 7\textsuperscript{th} on the Nature Index\footnote{\url{https://www.nature.com/articles/d41586-021-03065-6}}, despite its population being low in contrast to the other countries from the top 10. In addition, growth in Canada Institutional Repositories (IRs) has been fast;~\cite{loan2020global} report that in 2020, Canada did not even get into the ranking of countries with the most IRs, meaning it had less than 57. At the time of writing, the current number of listed repositories on OpenDOAR for Canada is 99, including IRs from both universities and other institutions. Canada also has a unique bilingual culture which, as we show in Section~\ref{Sec:DataAna}, creates additional inconsistencies in metadata for ETDs. Understanding how Canadian UIRs are organizing ETDs is therefore important to ensure that UIRs grow toward a state in which they could be easily used by aggregators, researchers, and bibliominers. In general, as noted in~\citep{najjar2003actual}, the study of the use of repositories allows the development of better tools for sharing information and creating better standards. The last study of UIRs in Canada was in 2010, where~\cite{park2011metadata} showed inconsistencies in the use of metadata on a subset of 10 Canadian universities. Amid Canada's evolving UIRs, we aim to see if the situation has changed and, if so, in which direction.


In this article, we conduct an analysis of all Canadian UIRs. Out of 106 universities, we find 57 repositories that we could extract the content of (we found 62 Canadian UIRs in total, with 5 returning errors during the extraction), and present a study of their main characteristics, commonalities, and differences. We find problems in conformity, not only in the language used for describing elements (e.g.: \xmlval{thèse} vs. \xmlval{thesis}), but also in the use of metadata elements, such as dates in the \xmltag{description} element. In this work, we focus our analysis on the use of the OAI\_DC metadata format, but we also provide a complete list of the metadata formats supported by each UIRs.

The article is structured as such. In Section~\ref{sec:preliminaries}, we provide a definition of the key concepts essential to this paper, such as IRs, UIRs, and metadata, followed by a literature review in Section~\ref{Sec:LittRev}. Sections~\ref{Sec:Repo} and~\ref{sec:metadata_formats} discuss the collection process of UIRs as well as the metadata formats that are used in Canadian UIRs. Then in Sections~\ref{Sec:DataColl} and~\ref{Sec:DataAna} we describe our collection effort and analyse the resulting metadata dataset. Finally, we use this knowledge to set forth recommendations in Section~\ref{Sec:Recomm} and conclude in Section~\ref{Sec:Conclusion}.

\section{Preliminaries}
\label{sec:preliminaries}

We begin by providing an overview of key concepts central to this paper, namely IRs, metadata, metadata formats, and methods for accessing them. IRs, or Institutional Repositories, are essentially collections of documents hosted by institutions. Our focus in this study is on University Institutional Repositories (UIRs), which are accessible online through the OAI-PMH protocol (Open Archives Initiative Protocol for Metadata Harvesting).

As its name implies, this protocol allows the collection and sharing of metadata (data describing aspects of a document, such as the publication date, the title, or a summary) using common commands. To do so, the first thing needed is an access point to the repository. For example, Université de Montréal UIR is hosted at \url{https://papyrus.bib.umontreal.ca/xmlui/}, and the OAI-PMH access point is \url{https://papyrus.bib.umontreal.ca/oai/}. As we show in Section~\ref{Sec:Repo}, the URL structure is not regulated or consistent and is often surprisingly difficult to find. From the access point one can, in accordance to the protocol, use additional verbs to gain access to the metadata. In this case, one could use the \textit{Identify} command (\url{https://papyrus.bib.umontreal.ca/oai/request?verb=Identify}) to get information about the repository. 



This paper focuses on metadata access, specifically examining three aspects: metadata formats, software utilized for OAI-PMH metadata management, and the metadata content itself.

When accessing the metadata, users must specify the desired output format. The metadata is always presented in XML, which is a markup language that employs tags to structure data. However, the chosen format determines the specific XML tags employed to convey information. In Section~\ref{sec:metadata_formats}, we provide detailed descriptions of the various formats utilized.

The Dublin Core Metadata Element Set (usually shortened as Dublin Core) is a compilation and description of 15 metadata elements (formally and internationally standardized as ISO 15836). It was developed by the non-profit organization \textit{Association for Information Science and Technology} within the Dublin Core Metadata Initiative project. Its goal is to define a minimal inventory of metadata necessary to record digital or physical resources of textual, visual, or auditory nature. It enables the documentation science community to have a common, freely-available and inter-operable specification, as well as a shared metadata vocabulary to describe resources of diverse nature.

It is distinct from the OAI\_DC format (also informally called Dublin Core) which is an implementation of the metadata elements described by the Dublin Core Metadata Element Set. It formalizes the abstract metadata concepts into XML metadata tags as described in Section \ref{sec:metadata_formats}. We make in this work the distinction between the two concepts by using the term Dublin Core to refer to the Dublin Core Metadata Element Set and OAI\_DC to refer to the OAI\_DC format.

\section{Literature Review}
\label{Sec:LittRev}
The management of an OAI-PMH repository is a subject of interest due to its direct impact on the sharing and accessibility of knowledge. In this section, we provide a literature survey focusing on the conformity of IRs as well as the studies of national IRs across the world. For a broader literature review, we refer to~\cite{nisa2021systematic}.

Metadata conformity has been extensively studied in the past.~\cite{bueno2009study} take a special interest in the dissemination of \textit{learning objects} through OAI-PMH, analyzing 47 repositories listed on OpenDOAR and reporting inconsistencies in their metadata structures, effectively rendering those IRs inefficient for sharing knowledge.~\cite{efron2007metadata} examines the OAI-DC format of 23 repositories from DSPACE (a software used for creating and maintaining IRs) and also conclude that there are inconsistencies in the usage of metadata, with a large portion being left empty and with some repositories returning errors when accessing them.  Similarly,~\cite{ward2003quantitative} observes the OAI-DC of 100 data providers, finding that about half the tags of OAI-DC were used less than half the time. 

This conclusion is shared also by~\cite{shreeves2003harvesting}, which however add that the general use of metadata attributes varies depending on the communities (museum, academic, or digital libraries). By examining several repositories,~\cite{dushay2003analyzing} conclude that they can classify inconsistencies into missing data, confusing data, incorrect data, and insufficient data.~\cite{shreeves2005quality} establish ways to evaluate the data from repositories, noting that conformity to norms is important for aggregators. They evaluate some databases along different axes (completeness, consistency, and ambiguity), concluding that a lot of the data is not suitable for aggregation. More general reflections on the use and management of metadata have also been reported in~\cite{barton2003building, schopfel2014open}.

The idea to explore and analyze a country's IRs to assess their quality is not novel, as it is an important notion to assess the accessibility of the produced outputs. In fact, countries face specific cultural challenges which make individual assessment excessively important. For example,~\cite{elahi2018open} study the state of IRs and OAI in Bangladesh, a country with a fairly low literacy rate (61.5\% at the time of the study, currently 74.7\%\footnote{\url{https://www.thedailystar.net/youth/education/news/bangladeshs-literacy-rate-now-7466-3080701}}) and find that despite several obstacles such as lack of openness or the presence of predatory repositories, the use of IRs was steadily growing and being centralized.~\cite{ukwoma2017institutional} review the state of Nigerian IR, concluding that more funding, conferences, and communications about the issue is necessary to assure a good use of IRs.

\cite{rodriguez2007science, abadal2010open} explore the state of IR in Spain, noting that it stands behind its neighbouring countries, but also observe that there is significant growth in IRs, giving hope for the future. Similar studies have been conducted in India~\cite{singh2016open}, Nigeria~\cite{christian2009issues}, Zimbabwe~\citep{kusekwa2014open}, and Bangladesh~\citep{islam2013institutional}. 

While most studies find things severely needing improvement, with both a lack of participation and adhesion to OAI-PMH standards from universities, in some cases findings are positive. Noticeably,~\cite{shin2010challenges} find the governmental repositories in Korea to be well-built, although they note that more involvement from researchers and universities is required. 

As we can see, the exploration of IRs from national bodies, and most notably Universities, can reveal important trends in how effectively the country's research may be disseminated. Most studies find the metadata to be severely lacking, but this is not always the case. It is also interesting to note that the state of IRs is often linked to the geopolitical situation of the country, and Canada's, with its strong research output but also its bilingual culture, proves to be an interesting subject for metadata analysis.

\section{Repositories}
\label{Sec:Repo}

We first need to find the address of the repository for each university, and its OAI-PMH access point. This step is complex for three reasons: 1. There is no guarantee that a given university has such a repository, 2. OAI-PMH access points can be difficult to find from the repository URL\footnote{The OAI-PMH URL for a repository is typically found in subdirectories of the main URL. While some patterns are common (e.g., \texttt{/oai}, \texttt{/oai2}), other directory structures are unique and not easily guessable, such as \texttt{/server/oai}.}, and 3. the information available online about the OAI-PMH URLs is sparse and lacking. The main resource that lists IRs with OAI-PMH enabled is OpenDOAR\footnote{\url{https://v2.sherpa.ac.uk/view/repository_by_country/Canada.html}}, but similarly to~\cite{islam2013institutional}, we found that many UIRs were missing from the list or had incorrect information (out of the initial UIRs we found, 23 were not listed on OpenDOAR or had incorrect information). Another existing resource is BorealisData\footnote{\url{https://borealisdata.ca}}, which stores theses and documents from universities, but it does not list the URLs for the OAI-PMH access points, and we found that the data listed is severely lacking. For example, it reports only 1425 documents for the UIR of University of Montréal, in contrast with the 27040 documents we collected.

Using a combination of the resources described above, independent search, and directly contacting universities, we found 62 repository addresses, among which five were not accessible for data collection (either timed out or returned errors while trying to harvest the data). We list all collected URLs in Appendix~\ref{App:RepUrls}, which to our knowledge is the most exhaustive list of Canadian UIRs to date. 

An important thing to note is the software used for the creation and management of the UIRs.  This software not only impacts which metadata formats are available, but is also central to the set up of the UIRs. Most studies found that a majority of IRs used DSpace~\citep{loan2020global}, but interestingly, we found that the proportion of DSpace repositories in Canadian UIRs is smaller. We believe that this is due to the large presence of Islandora, which is a Canadian Open Source software for setting up and managing IRs. A visualization of the softwares used in the UIRs we collected is presented in Figure~\ref{fig:DistributionProgs} and the complete list can be found in Appendix~\ref{App:RepUrls}.

\begin{figure}
    \centering
    \includegraphics[scale=0.5]{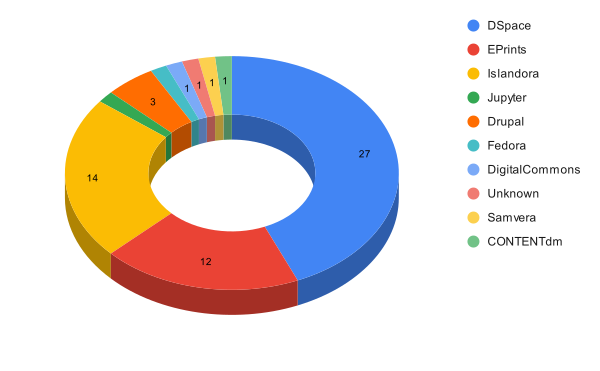}
    \caption{Distribution of the programs used for managing the 62 IRs we found the OAI-PMH access point of.}
    \label{fig:DistributionProgs}
\end{figure} \vspace{-5pt}

\section{Metadata formats}
\label{sec:metadata_formats}
In this section, we explore available metadata formats in ETDs repositories. We discovered 35 different metadata prefixes, but no standardized naming conventions. This results in multiple instances of various prefixes describing the same format (e.g. the ETDMS format is often provided under the prefixes \emph{etdms}, \emph{oai_etdms}, \emph{etd-ms}, etc.). Some repositories provide multiple prefix aliases for some common metadata formats, and some providers maintain multiple competing versions of the same metadata format (e.g. \emph{etdms10} vs \emph{etdms11} or \emph{marc} vs \emph{marc21}). After removing aliases, versions, and outlier formats with minimal use, we identify 10 core metadata formats: OAI\_DC, QDC, ETDMS, DIDL, MARC, METS, MODS, ORE, RDF, and UKETD\_DC. We also excluded the DIM and XOAI formats as they seem to be provided only by DSpace, and we could find little to no documentation on them.


In this section, we briefly present the most commonly used metadata formats. Analysis of those have been done in the past~\citep{burnett1999comparison}, but were not specific to UIRs, and we found wide differences between the metadata formats considered in this work and the ones provided by the UIRs we collected.

\textbf{OAI\_DC} (Open Archives Initiative--Dublin Core): As per the OAI-PMH specifications, OAI\_DC is the only required metadata format for all OAI-PMH repositories. OAI\_DC is based on the Dublin Core Metadata Element\footnote{\url{https://www.dublincore.org/specifications/dublin-core/}} Set which is composed of 15 metadata elements: \xmltag{Contributor}, \xmltag{Coverage}, \xmltag{Creator}, \xmltag{Date}, \xmltag{Description}, \xmltag{Format}, \xmltag{Identifier}, \xmltag{Language}, \xmltag{Publisher}, \xmltag{Relation}, \xmltag{Rights}, \xmltag{Source}, \xmltag{Subject}, \xmltag{Title}, and \xmltag{Type}. An example of a simplified metadata record in OAI\_DC can be found in Figure~\ref{fig:XMLStructure}.

\textbf{QDC} (Qualified Dublin Core) is a modified Dublin Core schema that adds three metadata elements: \xmltag{Audience}, \xmltag{Provenance}, and \xmltag{Rights Holder}. Furthermore, QDC specifies a set of qualifiers that can be applied to various elements to either further refine the function of an element (e.g. the qualifier \emph{abstract} to the element \xmltag{Description}) or specify the encoding scheme used for an element. When using OAI-PMH, the most frequent metadata prefixes for QDC are \emph{qdc}, \emph{dqc} or \emph{oai\_qdc}. Forty of the repositories provide QDC.

\textbf{ETDMS} (Electronic Thesis and Dissertation Metadata Standard\footnote{\url{https://ndltd.org/wp-content/uploads/2021/04/etd-ms-v1.1.html}}) is a modification to the OAI\_DC standard explicitly made for ETDs. On top of the 15 elements already present in OAI\_DC, ETDMS adds a new \xmltag{Degree} tag containing one or more subtags among \xmltag{Name}, \xmltag{Level}, \xmltag{Discipline}, and \xmltag{Grantor}. ETDMS also provides some qualifiers for the existing elements, such as a \emph{role} qualifier for the \xmltag{Description} element, and attempts to normalize the content of some elements (e.g. only 0, 1 and 2 are supposed to be valid codes for the \xmltag{Rights} element). The OAI-PMH metadata prefix for ETDMS is most often either \emph{etdms} or \emph{oai\_etdms}. Out of the 62 UIRs used in this study, 49 provide the ETDMS format.

\textbf{DIDL} (Digital Item Declaration Language\footnote{\url{https://www.xml.com/pub/a/2001/05/30/didl.html}}) was established as a mean to be able to describe raw content such as music or images, and is intended as a very general mean to describe content. DIDL is supplied by 38 repositories out of the 62 queried UIRs.


\begin{figure}
    \centering
\begin{lstlisting}[language=CustomXML]
<metadata>
<oai_dc:dc xsi:schemaLocation="http://www.openarchives.org/OAI/2.0/oai_dc/ http://www.openarchives.org/OAI/2.0/oai_dc.xsd http://www.w3.org/1999/02/22-rdf-syntax-ns# http://www.openarchives.org/OAI/2.0/rdf.xsd ">
<dc:title>Personality extraction through LinkedIn</dc:title>
<dc:creator>Piedboeuf, Frédéric</dc:creator>
<dc:contributor>Langlais, Philippe</dc:contributor>
<dc:contributor>Lapalme, Guy</dc:contributor>
<dc:subject>Extraction de personalité</dc:subject>
<dc:subject>MBTI</dc:subject>
<dc:subject>DiSC</dc:subject>
<dc:subject>LinkedIn</dc:subject>
<dc:subject>Réseau sociaux</dc:subject>
<dc:subject>Profilage d'auteur</dc:subject>
<dc:subject>Personality Extraction</dc:subject>
<dc:subject>Social Network</dc:subject>
<dc:subject>Author Profiling</dc:subject>
<dc:subject>
Applied Sciences - Computer Science / Sciences appliqués et technologie - Informatique (UMI : 0984)
</dc:subject>
<dc:description>
L'extraction de personnalité sur les réseaux sociaux est un domaine qui n'a que récemment commencé à capturer l'attention des chercheurs. La tâche consiste à, en partant d'un corpus de profils d'utilisateurs de réseaux sociaux, être capable de classifier leur personnalité correctement, selon un modèle de personnalité tel que défini en psychologie. Ce mémoire apporte trois innovations au domaine. Premièrement, la collecte d'un corpus d'utilisateurs LinkedIn. Deuxièmement, l'extraction sur deux modèles de personnalités, MBTI et DiSC, l'extraction sur DiSC n'ayant pas encore été faite dans le domaine, et finalement, la possibilité de passer d'un modèle de personnalité à l'autre est explorée, dans l'idée qu'il serait ainsi possible d'obtenir les résultats de multiples modèles de personnalités en partant d'un seul test.
</dc:description>
<dc:date>2019-11-19T19:23:16Z</dc:date>
<dc:date>NO_RESTRICTION</dc:date>
<dc:date>2019-11-19T19:23:16Z</dc:date>
<dc:date>2019-10-30</dc:date>
<dc:date>2019-05</dc:date>
<dc:type rdf:resource="http://purl.org/coar/resource_type/c_46ec" xml:lang="en">thesis</dc:type>
<dc:type rdf:resource="http://purl.org/coar/resource_type/c_46ec" xml:lang="fr">thèse</dc:type>
<dc:identifier>http://hdl.handle.net/1866/22536</dc:identifier>
<dc:language>eng</dc:language>
<dc:format>application/pdf</dc:format>
</oai_dc:dc>
</metadata>

\end{lstlisting}
    \caption{A simplified example of an OAI-PMH record (in Dublin Core format), from the repository Papyrus, of the University of Montreal. Notably, we exclude the header which is returned along the metadata elements while querying the OAI-PMH repository.}
    \label{fig:XMLStructure}
\end{figure} 

\textbf{MARC} (MAchine Readable Cataloging\footnote{\url{https://www.loc.gov/marc/}}) is a legacy way to represent bibliographic metadata using a set of numeric codes (for example, the code 245 is used to define a title entry), which incidentally makes MARC one of the less human-readable format presented here. MARC is found in 26 repositories. 

\textbf{METS} (Metadata Encoding and Transmission Standard\footnote{\url{https://www.loc.gov/standards/mets/}}) is maintained by the Library of Congress and was designed to be a wrapper for other metadata formats such as MARC. In essence, while Dublin Core and MODS are intended for cataloguing and giving high level descriptions of a document, METS is intended to give a full description of the object, referencing its structure. Similarly to OAI\_DC, METS is intended to structure the XML document, and recommends the use of MODS and Dublin Core for metadata elements, among others~\citep{cantara2005mets}. METS is provided by 37 of the queried repositories. 

\textbf{MODS} (Metadata Object Description Schema\footnote{\url{https://www.loc.gov/standards/mods/}}) is also maintained by the Library of Congress and is a schema intended for various purposes. It proposes 10 top-level tags and 105 sublevel tags, and is intended to be an overlay on the MARC format to convert the MARC schema into more human-readable tags. We found that 42 of our queried repositories supply this format.

\textbf{ORE} (Object Reuse and Exchange) facilitates aggregations of Web resources, like image collections or HTML documents with interconnected webpages. We found ORE in 26 repositories.

\textbf{RDF} (Resource Description Framework) is a very general structure of XML elements. In fact, all other schemas are built on RDF. By observing the RDF data of universities supporting it, we found that it was used as a wrapper for OAI\_DC tags, making it not only superfluous but confusing for users. RDF was found in 37 of the repositories.

\textbf{UKETD\_DC} is a format developed by EThOS (Electronic Thesis Online Service) for describing UK theses. Very little information can be found about it and in fact, we could find no official documentation for it. This format is provided by 37 of the queried UIRs.

\section{Data collection and cleaning}
\label{Sec:DataColl}

We now aim to analyse how metadata is presented in the UIRs. We extract all data from all universities using the \texttt{pyoai} python package\footnote{\url{https://github.com/infrae/pyoai}}, which allows the extraction of multiple types of schemas. Because OAI\_DC is the only format that is available through all repositories, we focus on the extraction of this format, even if it is not the most appropriate format for ETDs\footnote{The code is available at \url{https://github.com/dahrs/oai-pmh_canadian_universities}}.

Metadata for a given document is organized in an XML structure, as shown in Figure~\ref{fig:XMLStructure}, but using \texttt{pyoai} allows us to extract automatically the content of the XML file in a more readable format. It is important to note that for the extraction to work correctly, names and document structure need to be standardized. For instance, if a university writes the \xmltag{date} element as ``\emph{Dote}'' (with a typo), the parser will not catch it. Given that the conversion of metadata to XML is generally handled by third party software, the chance of erroneous naming is quite low and in fact when viewing the data we found no such instances. 


\begin{table}
    \centering
    \caption{Percentage of empty entries for each element of the OAI\_DC extracted documents.}
    \begin{tabular}{l@{\hskip 20px}r}
        \hline
        \textbf{Tag} & \textbf{Frequency of absence}\\
        \hline
        Source & \percentbar{0.99}{100pt}{99.4\%} \\
        Coverage & \percentbar{0.95}{100pt}{95.0\%}\\
        Relation & \percentbar{0.74}{100pt}{74.4\%}\\
        Contributor & \percentbar{0.60}{100pt}{60.4\%}\\
        Rights & \percentbar{0.55}{100pt}{55.4\%}\\
        Publisher & \percentbar{0.49}{100pt}{49.2\%}\\
        Subject & \percentbar{0.33}{100pt}{32.9\%}\\
        Format & \percentbar{0.30}{100pt}{29.5\%}\\
        Language & \percentbar{0.23}{100pt}{23.1\%}\\
        Abstract & \percentbar{0.18}{100pt}{17.7\%}\\
        Creator & \percentbar{0.13}{100pt}{12.6\%}\\
        Type & \percentbar{0.07}{100pt}{7.0\%}\\
        Date & \percentbar{0.03}{100pt}{3.0\%}\\
        Identifier & \percentbar{0.03}{100pt}{0.3\%}\\
        Title & \percentbar{0.0}{100pt}{0.0005\%}\\
        \hline
    \end{tabular}
    \label{tab:percentEmpty}
\end{table}

Extraction over the 57 repositories yielded 728,754 documents. Initially, we examine the percentage of empty elements in the data collected, providing insights into UIRs' prioritization and perceived usefulness of specific elements. Table~\ref{tab:percentEmpty} presents the results, revealing that some elements are largely neglected in the Canadian UIRs, particularly \xmltag{Source} and \xmltag{Coverage} (refer to Section~\ref{Sec:DataAna} for element descriptions).

\cite{shreeves2005quality} focus on developing efficient tools for assessing metadata and note that to be complete at least 8 tags should be present: \xmltag{Title}, \xmltag{Creator}, \xmltag{Subject}, \xmltag{Description}, \xmltag{Date}, \xmltag{Format}, \xmltag{Identifier}, and \xmltag{Rights}. We found that through our collections, only 17.5\% of the extracted documents filled all these tags. This does not indicate that data is absent, as the relevant information may be present in another tag. For example, keyphrases occasionally appear at the end of the abstract in the \xmltag{description} tag instead of inside the \xmltag{subject} tag.


\section{Data analysis}
\label{Sec:DataAna}

In this section, we discuss how tags are used in the UIRs we collected. We perform a qualitative evaluation using representative samples from each university and include quantitative evaluations when needed.

\textbf{Contributor:} This tag is intended to specify the co-authorship, supervisor participation, affiliation to a laboratory, etc. While it is absent in more than half of the entries, those that use this tag tend to input the correct information for it. In Figure~\ref{fig:XMLStructure} the two contributors reference the co-supervisor of the master thesis.

\textbf{Coverage: } According to guidelines set up by OAI\_DC, and followed by the ETDMS format, \xmltag{coverage} is expected to be used to represent the legal, spatial, or temporal coverage of the subject. While not used extensively, we found that in most cases, it is used for the geographical coverage. However, even within that use there is no coherence, with some ETDs using place names (e.g. \xmlval{Fidji}), and some using coordinates (e.g. \xmlval{49.13, -122.871}). When used for temporal coverage, dates are, as we describe in more details below, not standards. Furthermore, a proportion of the data in the \xmltag{coverage} tags represents nothing, such as \xmlval{----------}. Ultimately, it would be very hard for aggregators at the moment to use this tag for analysis due to not only the various information that this field could contain but also the lack of convention. 

\textbf{Creator}: In the same vein, \xmltag{creator} is a tag that represents any entity responsible for creating the resource. Most often this matches the author(s), but in some cases it is used to divulge institutions (e.g.: \xmlval{Ontario Agricultural College}). We found that overall this tag seems to be used appropriately, probably due to the high information value of the author names, which universities are deeply invested in cataloguing correctly. Our example in Figure~\ref{fig:XMLStructure} follows the convention of listing the author's name in the \xmltag{Creator} tag.


\textbf{Date:} is another vital tag for data analysis. As is common practice in librarianship and information science, most dates follow ISO 8601 formats (e.g. \xmlval{2019-11-19T19:23:16Z} or \xmlval{2019-05}), as shown in Appendix~\ref{app:date}. However, there are exceptions such as the American anglophone standard (month-day-year), or even non-date entries such as \xmlval{NO\_RESTRICTION} (see Figure~\ref{fig:XMLStructure}). Additionally, it is not uncommon to encounter multiple dates for a single document without clear indications of their specific meanings (submission date, presentation date, graduation date, etc.), as exemplified in Figure~\ref{fig:XMLStructure}.

\textbf{Description:} Within the context of ETDs, the \xmltag{description} element is most often used for the abstract of the document. However, some universities also provide other information in that field, such as information about the origin of the document and even, in some cases, dates. When multiple abstracts are provided (e.g. when the abstract is provided in multiple languages), these abstracts are sometimes concatenated together in a single \xmltag{Description} element. There are also cases of single abstracts being split between multiple \emph{Description} elements (e.g. one line per element). \xmltag{description} is one of the most high value elements of ETDs, and the bad formatting is certainly a heavy obstacle to its employment by aggregators or analysts. Our running example presents the French abstract of the thesis as the description element. 


\textbf{Format:} The \xmltag{format} element mostly contains references to the file type of the document the metatada references. Given that UIRs contain mostly ETDs, most records thus contain some variation of \xmlval{pdf} (the most common being \xmlval{application/pdf}, as in Figure~\ref{fig:XMLStructure}).

\textbf{Identifier:} This field is used to give a link to the underlying document. These can either be direct links to the IR or handles such as Handle.Net or DOI. Our example in Figure~\ref{fig:XMLStructure} gives a link to the document in the OAI-PMH IR.

\textbf{Language:} Since the OAI-PMH protocol only allows the extraction of the metadata and not the document itself, it is quite difficult to analyse if the \xmltag{Language} tag is correctly assigned. However, as for many other elements, we notice a lack of uniformity between UIRs. We find different naming conventions across the UIRs such as fully spelled out names (e.g.: \xmlval{english}, \xmlval{french}, \xmlval{français}, etc.) as well as various ISO 639 formats (\xmlval{en}, \xmlval{fra}, \xmlval{en\_us}, etc.). It is not surprising to note that across the chronology (see Figure~\ref{fig:langyear}), records are mostly declared to be in either English or in French. Table~\ref{tab:lang} shows the distribution of assigned languages among the extracted records. The regular expressions used for normalizing the languages are shown in Appendix~\ref{app:lang_regexes}. 

\textbf{Publisher:} As noted in Table~\ref{tab:percentEmpty}, most entries do not contain a \xmltag{publisher} tag. This is normal, since a lot of theses produced at universities do not go through the usual publisher pipeline but are simply uploaded online upon acceptance, as is reflected in our example. Still, this convention is not adopted universally. For example, we found almost 13K documents which were tagged as \xmlval{article} but had no publishers. Some entries also had the \xmlval{These} type (in the \xmltag{Type} tag) and the university as a publisher. This, ultimately, makes the use of the \xmltag{Publisher} tag difficult for bibliomining or aggregation.

\textbf{Relation:} is generally of little use for ETDs, as it is supposed to represent a "related resource". In fact, this and the \xmltag{Source} tags are the only two tags for which the ETDMS guidelines give no recommendations of what to input. Still, some uses of the \xmltag{relation} tag we could see were, among others, a link pointing to the specific ETD in the UIR, books or chapter names (such as \xmlval{DMS-676-IR}), or what seems like general themes \xmlval{Education \& culture}. Ultimately, this tag may be useful for searching information, but the lack of guidelines to use it hinders aggregation and data analysis. We note that in our example in Figure~\ref{fig:XMLStructure}, the relation tag is also absent.

\textbf{Rights:} Approximately half of the theses appearing in all repositories do not contain any mention to the rights associated with the document. One of the reasons for this omission is the fact that for theses and institutional documents, the rights of property and ownership may be derived from the author(s) of the theses and set on the year of publication; two concepts that already appear in the record.

\begin{table}
    \centering
        \caption{Language distribution according to the provided \xmltag{Language} tag. The English category regroups all records labelled \xmlval{english}, \xmlval{anglais}, \xmlval{en}, \xmlval{eng}, \xmlval{en\_us}, or \xmlval{en\_ca} and the French category is composed of all the records labelled either \xmlval{french}, \xmlval{français}, \xmlval{francais}, \xmlval{fre}, \xmlval{fra}, or \xmlval{fr}.}
    \begin{tabular}{l@{\hskip 10px}r}
        \hline
        \textbf{Languages}&\textbf{Frequency}\\
        \hline
        English&\percentbar{0.78}{100pt}{77.71\%}\\
        French&\percentbar{0.18}{100pt}{18.24\%}\\
        Other&\percentbar{0.04}{100pt}{4.05\%}\\
        \hline
    \end{tabular}
    \label{tab:lang}
\end{table}

\begin{table}
    \centering
    \begin{minipage}[t]{0.45\textwidth}
        \centering
        \caption{Most frequent content of the \xmltag{Type} tag as presented in the extracted metadata.}
        \begin{tabular}{l@{\hskip 40px}r}
            \hline
            \textbf{Types}&\textbf{Frequency}\\
            \hline
            Text&20.91\%\\
            Thesis&15.82\%\\
            Dataset&13.46\%\\
            Image&12.17\%\\
            Article&9.24\%\\
            Journal contribution&9.22\%\\
            Figure&8.12\%\\
            Master thesis&5.10\%\\
            Nonpeerreviewed&3.28\%\\
            Thèse&2.86\%\\
            Journal article&2.81\%\\
            Online resource&2.31\%\\
            \hline
        \end{tabular}
        \label{tab:type_frequency}
    \end{minipage}\hfill
    \begin{minipage}[t]{0.45\textwidth}
        \centering
        \caption{Most frequent types grouped by value categories, using the regular expressions presented in Appendix~\ref{app:regexes}.}
        \begin{tabular}{l@{\hskip 75px}r}
            \hline
            \textbf{Types}&\textbf{Frequency}\\
            \hline
            Thesis&30.70\%\\
            Media&16.15\%\\
            Article&13.51\%\\
            Dataset&13.46\%\\
            Book&1.28\%\\
            Presentation&1.04\%\\
            Poster&0.33\%\\
            \hline
        \end{tabular}
        \label{tab:regex_type_frequency}
    \end{minipage}
\end{table}

\textbf{Source:} among all tags in the OAI\_DC format, this tag is the most absent overall, and our example in Figure~\ref{fig:XMLStructure} is no exception. We believe this is partly due to the fact that theses do not always have a clear source (for example the URL, university name, financing institution, etc. would all be valid sources) and also because the utility of the \xmltag{Source} tag is higher for documents of a different nature such as maps, media files or images.

\textbf{Subject:} The \xmltag{Subject} element is intended to contain the keyphrases for the document. However, the subject's link to the thesis sometimes seems quite distant. This is due to the policy of some universities to automatically generate (often generic) keyphrases. Furthermore, it is not rare to see keyphrases that have been provided by the users with incorrect separators, thus preventing an automatic system from correctly identifying them as distinct keyphrases. This results in several subject elements containing keyphrases of the form \xmlval{A;B;C} for example where \xmlval{A}, \xmlval{B}, and \xmlval{C} should have been instead identified as 3 separate entries. In our example, we can see both French and English keyphrases being present in the \xmltag{Subject} tag.


\textbf{Title:} As noted in Table~\ref{tab:percentEmpty}, entries for the \xmltag{title} tag are rarely empty. Although some other tags might be ambiguous and allow for multiple interpretations, in the case of theses this tag is straightforward, and we can see the classical use in Figure~\ref{fig:XMLStructure}.


\textbf{Type:} One of the most present tags, \xmltag{type} is potentially also one of the most useful ones for aggregators and for analysis purposes. Here again, there is a lack of uniformity between records of different universities. Our example provides a demonstration of this, with both the \xmlval{thesis} and \xmlval{thèse} information provided. Table~\ref{tab:type_frequency} presents a list of the most frequent types. In this form, it is difficult to extract any meaningful statistic. We therefore use regular expressions to group some types into more meaningful categories (presented in Appendix~\ref{app:regexes}), and the result is shown in Table~\ref{tab:regex_type_frequency}.

\begin{table}
    \centering
    \caption{Amount of documents published per decade, all repositories combined. At the moment of the writing, the 2020 decade is not over (2020-2023), making the numbers from this decade not comparable to the previous.}
    \begin{tabular}{l@{\hskip 20px}r}
        \hline
        \textbf{Year}&\textbf{Documents}\\
        \hline
        1920 & \percentbar{0.001}{100pt}{500}\\
        1930 & \percentbar{0.001}{100pt}{571}\\
        1940 & \percentbar{0.001}{100pt}{485}\\
        1950 & \percentbar{0.002}{100pt}{882}\\
        1960 & \percentbar{0.003}{100pt}{1745}\\
        1970 & \percentbar{0.026}{100pt}{11476}\\
        1980 & \percentbar{0.032}{100pt}{14235}\\
        1990 & \percentbar{0.048}{100pt}{21008}\\
        2000 & \percentbar{0.263}{100pt}{115014}\\
        2010 & \percentbar{1.00}{100pt}{437199}\\
        \grey{2020} & \percentbar{0.243}{100pt}{\grey{106621}}\\
        \hline
    \end{tabular}
    \label{fig:percentEmpty}
\end{table}

Our qualitative analysis identifies three categories of metadata problems. Firstly, there are unused tags, such as the absence of a \xmltag{publisher} tag for articles or books. In our view, this is the least significant issue among the three, as missing information is easier to handle in analysis and research compared to incorrect information.

The second category is the fact that some information is stored in the wrong tag, such as the common use of the \xmltag{description} tag used for storing the date, instead of the \xmltag{date} tag. This is more problematic, even if in some cases it can be mitigated. For example, one could attempt to recognize whether the description is a date or not, by checking if it corresponds to the ISO norm. That technique, however, would not eliminate completely the problem, and would still demand considerable efforts in order to obtain clean data.
\pgfplotsset{
        /pgfplots/bar shift auto/.style={
            /pgf/bar shift={
                -0.5*(\numplotsofactualtype/2*\pgfplotbarwidth + ((\numplotsofactualtype/2)-1)*(#1))
            },
        },
    }
\begin{figure}[b]
        \centering
        \scalebox{0.75}{
        \begin{tikzpicture}
            \begin{axis}[
                    ybar,
                    bar width=.25cm,
                    width=\textwidth,
                    height=.4\textwidth,
                    legend style={at={(0.2,1)},
                        anchor=north,legend columns=-1},
                    symbolic x coords={2000, 2001, 2002, 2003, 2004, 2005, 2006, 2007, 2008, 2009, 2010, 2011, 2012, 2013, 2014, 2015, 2016, 2017, 2018, 2019, 2020, 2021, 2022, 2023},
                    xtick=data,
                    x tick label style={rotate=45,anchor=east},
                    ymin=0,
                    ymax=1,
                    xlabel={Year},
                    xlabel style={at={(0.5,-0.1)}},
                    ylabel={Ratio},
                ]
\addplot +[ybar, color=red!40!white, draw=red] plot coordinates {(2000, 1.0)(2001, 1.0)(2002, 1.0)(2003, 1.0)(2004, 1.0)(2005, 1.0)(2006, 1.0)(2007, 1.0)(2008, 1.0)(2009, 1.0)(2010, 1.0)(2011, 1.0)(2012, 1.0)(2013, 1.0)(2014, 1.0)(2015, 1.0)(2016, 1.0)(2017, 1.0)(2018, 1.0)(2019, 1.0)(2020, 1.0)(2021, 1.0)(2022, 1.0)(2023, 1.0)};
\addplot +[ybar, color=blue!40!white, draw=blue] plot coordinates {(2000, 0.43200000000000005)(2001, 0.41100000000000003)(2002, 0.399)(2003, 0.352)(2004, 0.22999999999999998)(2005, 0.08099999999999996)(2006, 0.18700000000000006)(2007, 0.526)(2008, 0.22599999999999998)(2009, 0.19399999999999995)(2010, 0.21499999999999997)(2011, 0.23399999999999999)(2012, 0.20599999999999996)(2013, 0.11199999999999999)(2014, 0.121)(2015, 0.138)(2016, 0.21399999999999997)(2017, 0.18000000000000005)(2018, 0.30100000000000005)(2019, 0.133)(2020, 0.14600000000000002)(2021, 0.17900000000000005)(2022, 0.16700000000000004)(2023, 0.14200000000000002)};
\addplot +[ybar, color=black!40!white, draw=black!80!white] plot coordinates {(2000, 0.010000000000000064)(2001, 0.009000000000000008)(2002, 0.008000000000000007)(2003, 0.011999999999999955)(2004, 0.010999999999999982)(2005, 0.004999999999999963)(2006, 0.004000000000000059)(2007, 0.379)(2008, 0.06999999999999998)(2009, 0.03799999999999995)(2010, 0.04499999999999996)(2011, 0.01999999999999999)(2012, 0.021999999999999964)(2013, 0.010999999999999982)(2014, 0.044)(2015, 0.034000000000000016)(2016, 0.011999999999999955)(2017, 0.006000000000000061)(2018, 0.010000000000000064)(2019, 0.0010000000000000009)(2020, 0.004000000000000031)(2021, 0.0020000000000000573)(2022, 0.005000000000000032)(2023, 0.0030000000000000027)};
\legend{en, fr, other}
            \end{axis}
        \end{tikzpicture}}
        \caption{Ratio of theses according to the declared language of the theses, from 2000 to 2023. Please note that at the moment of analysis, the 2023 year is not over (January-March).}
        \label{fig:langyear}
\end{figure}
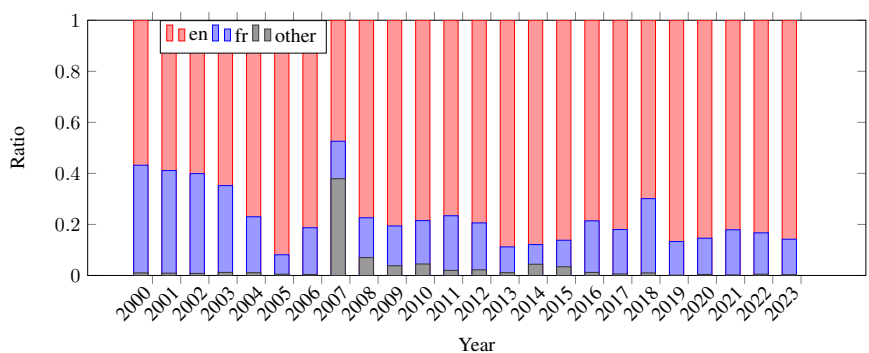

The last category of problems, which is in our opinion the biggest, is the lack of convention not only between universities, but internally as well. An example of this between universities would be the tagging of doctoral theses (in the \xmltag{type} tag) as \xmlval{Doctoral Thesis}, \xmlval{Thesis}, \xmlval{Thèse}, \xmlval{Thèse ou mémoire}, etc. An example of this internal dissonance is the fact that some departments will index documents with controlled vocabulary in the \xmltag{description} tag, such as the UMI or JEL standards, but not always. This, compounded to all tags and universities, makes it very difficult to obtain clean data either for analysis or aggregation.

\section{Recommendations}
\label{Sec:Recomm}
From our analysis of the repositories, we offer several recommendations that could greatly improve the state of institutional repositories in Canada, namely:

\textbf{Metadata schemes:} Firstly, we suggest eliminating non-essential schemes, like RDF, which are merely wrappers around OAI\_DC without added value. While more metadata schemes seems like a good idea, we argue that it makes little sense to have unused standards. Secondly, we advocate for a broader adoption of ETDMS, designed specifically for theses and dissertations. Furthermore, the institutional repositories that provide the ETDMS format should fully familiarize themselves with the metadata schema, use it to its full capacity, follow the defined standards, and keep up to date with updated versions.

\textbf{Naming convention:} We propose the adoption of naming conventions for some tags such as \xmltag{type} or \xmltag{format}. For example, we recommend a finite set of possible values (\xmlval{Thesis}, \xmlval{Article}, \xmlval{Book}, etc.) for the \xmltag{type} tag, which would facilitate the aggregation and identification of documents. Such conventions could be established by independent parties, the most logical representative in this case being Theses Canada.

\textbf{Language:} We encourage a wider use of the \emph{xml:lang} tag attribute to separately identify the language of each occurrence of the \xmltag{title}, \xmltag{description} (in particular for abstracts), and \xmltag{subject} tags. This language attribute should abide by the IETF's BCP 47 specification. We also encourage universities and providers to normalize the content of the \xmltag{language} describing the document language, instead of the various norms that have been used to specify the language until now. For example, while some use 2-letter (ISO 639-1) or 3-letter (ISO 639-3) descriptors, others use the language full name (e.g.: \xmlval{English}, \xmlval{French}, etc.), as shown in Section~\ref{Sec:DataAna}. In order to ease the extraction and language identification of records from multiple universities, we suggest that all institutions follow Theses Canada's recommendations and use the ISO 639-3's 3-letter code\footnote{\url{https://library-archives.canada.ca/eng/services/services-libraries/theses/pages/information-universities.aspx}}.

\textbf{Dates:} Dates should be formatted using any kind of ISO 8601 format for a more efficient parsing. The most commonly used metadata formats (including OAI\_DC) don't allow specifying what type of date (submission date, release date, etc.) is provided. Therefore, we recommend that only one date should be specified (by default: the release date, when the work is made public). The providers should also prevent the users from freely inputting dates and instead prefer a selection in a calendar. This would avoid erroneous dates like \xmlval{These are the annual proceedings of the Grand Lodge A.F. \& A.M. of Canada in the Province of Ontario covering (...)}.

\section{Conclusion}
\label{Sec:Conclusion}

Efficient metadata sharing is crucial for knowledge dissemination, aggregation, bibliomining, and analysis. Despite Canada's prominent position in international research and its unique cultural context, there is a lack of research on the state of metadata in Canadian UIRs.

In this paper, we address this issue by examining metadata in Canadian UIRs, highlighting their deficiencies in terms of rigour and uniformity. Our contributions include providing an up-to-date list of Canadian university UIRs with OAI-PMH access points, conducting qualitative and quantitative data analysis, and offering recommendations for improving metadata quality.

There is no easy solution to the metadata problem. As noted, user-generated metadata saves both time and money and therefore can be a valuable practice. However, even if authors have been found to generate quality metadata in experimental settings~\citep{greenberg2002author}, there is always a possibility of failure from the users due to several factors, such as unclear explanations or lack of examples~\citep{ed2003higher}. Additionally, evolving input mechanisms over time create inconsistencies in guidelines across repositories. Still, there is evidence that the repeated calls of researchers for better metadata standards do not fall on deaf ears, as recent years have seen some remediation and cleaning efforts in UIRs~\citep{thompson2019case, stein2017achieving}.

We hope that this research inspires the adoption of better metadata maintenance practices, not only among Canadian universities but worldwide. By improving existing UIRs through remediation and cleaning, it becomes possible to identify significant research trends, such as thesis languages and subjects of interest. This information is vital for understanding research focus, dissemination, and for informed decision-making by funding institutions.

\clearpage
\bibliographystyle{apalike}
\bibliography{biblio}

\newpage
\appendix

\section{Repository URLs}
\label{App:RepUrls}
\begin{center}

\begin{table}[!h]
    \caption{List of all OAI-PMH access points. Stars indicate repositories for which we couldn't extract the data due to errors returned when using the OAI-PMH}
    \centering
\rowcolors{1}{}{lightgray}
\scriptsize
\begin{tabular}{llr}
    \hline
    \textbf{Universities}&\textbf{OAI Repositories}&\textbf{Software}\\
    \hline
    Acadia University & https://scholar.acadiau.ca/oai2 & Islandora\\
    Athabasca University & https://dt.athabascau.ca/oai/request & DSpace \\
    Bishop's University* & https://eprints.ubishops.ca/cgi/oai2 & EPrints \\
    British Columbia Institute of Technology & https://circuit.bcit.ca/repository/oai2 & Drupal\\
    Brock University & https://dr.library.brocku.ca/oai/request &  DSpace \\
    Cape Breton University & https://cbufaces.cairnrepo.org/oai2 & Islandora\\
    Capilano University & https://capu.arcabc.ca/oai2/request & Islandora \\
    Carleton University & https://curve.carleton.ca/oai-pmh/all & Drupal \\
    Collège Militaire Royal du Canada & https://espace.rmc.ca/oai/request & DSpace \\
    Dalhousie University &https://dalspace.library.dal.ca/oai/request & DSpace\\
    Emily Carr University of Art and Design & https://ecuad.arcabc.ca/oai2 & Islandora \\
    HEC Montréal & https://reflexion.hec.ca/in/rest/oai & Unspecified\\
    Institut National de la Recherche Scientifique & https://espace.inrs.ca/cgi/oai2 & EPrints \\
    Kwantlen Polytechnic University & https://kora.kpu.ca/oai2 & Islandora \\
    Lakehead University & https://knowledgecommons.lakeheadu.ca/oai/request & DSpace \\
    MacEwan University & https://roam.macewan.ca/server/oai/request & Islandora \\
    McMaster University & https://macsphere.mcmaster.ca/oai/request & DSpace\\
    Memorial University of Newfoundland & https://research.library.mun.ca/cgi/oai2 & CONTENTdm\\
    Mount Royal University & https://mru.arcabc.ca/oai2 & Islandora \\
    Mount Saint Vincent University & https://ec.msvu.ca/server/oai/request & DSpace\\
    NSCAD University & https://nscad.cairnrepo.org/oai2 & Islandora\\
    National School of Public Administration & https://espace.enap.ca/cgi/oai2 &EPrints\\
    OCAD University & https://openresearch.ocadu.ca/cgi/oai2 & EPrints \\
    Ontario Tech University & https://ir.library.ontariotechu.ca/oai/request & DSpace \\
    Queen's University & https://qspace.library.queensu.ca/oai/request & DSpace\\
    Royal Roads University & https://viurrspace.ca/oai/request & DSpace\\
    Saint Mary's University & https://library2.smu.ca/oai/request & DSpace\\
    Simon Fraser University* & https://summit.sfu.ca/oai/request & Drupal\\
    St. Francis Xavier University & https://stfxscholar.cairnrepo.org/oai2 & Islandora \\
    TELUQ University & https://r-libre.teluq.ca/cgi/oai2 & EPrints\\
    Thompsom Rivers University & https://tru.arcabc.ca/oai2/request & Islandora\\
    Trinity Western University & https://twu.arcabc.ca/oai2/request  & Islandora\\
    University of Alberta & https://era.library.ualberta.ca/oai & Jupyter\\
    Université Concordia d'Edmonton & https://spectrum.library.concordia.ca/cgi/oai2 & EPrints \\
    University of Calgary & https://prism.ucalgary.ca/oai/request & DSpace \\
    University of Guelph & https://atrium.lib.uoguelph.ca/oai/request & DSpace \\
    University of Manitoba & https://mspace.lib.umanitoba.ca/oai/request & DSpace \\
    University of Northern British Columbia & https://unbc.arcabc.ca/oai2/request & Fedora \\
    University of Prince Edward Island & https://islandscholar.ca/oai2 & Islandora \\
    University of Quebec, Abitibi-Temiscamingue & https://depositum.uqat.ca/cgi/oai2 & EPrints\\
    University of Quebec at Chicoutimi & https://constellation.uqac.ca/cgi/oai2 & EPrints \\
    University of Quebec in Montreal & https://archipel.uqam.ca/cgi/oai2 & EPrints \\
    University of Quebec, Trois-Rivieres & https://depot-e.uqtr.ca/cgi/oai2 & EPrints \\
    University of Regina & https://ourspace.uregina.ca/oai/request & DSpace \\
    University of Saskatchewan & https://harvest.usask.ca/oai/request & DSpace\\
    University of Toronto & https://tspace.library.utoronto.ca/oai/request & DSpace\\
    University of Victoria & https://dspace.library.uvic.ca/oai/request & DSpace\\
    University of Waterloo & https://uwspace.uwaterloo.ca/oai/request & DSpace \\
    University of Winnipeg & https://winnspace.uwinnipeg.ca/oai/request & DSpace\\
    University of the Fraser Valley & https://ufv.arcabc.ca/oai2/request & Islandora \\
    Université Laurentienne & https://zone.biblio.laurentian.ca/oai/request & DSpace \\
    Université Laval & https://corpus.ulaval.ca/oai/request & DSpace\\
    Université McGill* & https://escholarship.mcgill.ca/catalog/oai/request & Samvera \\
    Université de Lethbridge* & https://opus.uleth.ca/oai/request & DSpace \\
    Université de Moncton & https://udmscholar.cairnrepo.org/en/oai2 & Islandora\\
    Université de Montréal & https://papyrus.bib.umontreal.ca/oai/request & DSpace \\
    Université de Sherbrooke & https://savoirs.usherbrooke.ca/oai/request & DSpace \\
    Université d’Ottawa & https://ruor.uottawa.ca/oai/request & DSpace \\
    York University & https://yorkspace.library.yorku.ca/oai/request & DSpace\\
    Western University* & https://ir.lib.uwo.ca/do/oai & DigitalCommons \\
    École Polytechnique de Montréal & https://publications.polymtl.ca/cgi/oai2 & EPrints\\
    École de Technologie Supérieure & https://espace.etsmtl.ca/cgi/oai2 & EPrints \\
    \hline
\end{tabular}
\end{table} \vspace{-5pt}
\end{center}

\newpage
\section{Supported metadata prefixes}
\label{App:SuppFormats}

\begin{table}[h]
    \caption{Supported formats for all extracted universities. Stars indicate repositories for which we couldn't extract the data due to errors returned when querying the IR, and empty entries are universities for which the OAI-PMH command to list metadata formats returned an error. }
    \centering
    \scalebox{1.02}{
    \centering
    \tiny
    \rowcolors{1}{}{lightgray}
    \begin{tabular}{lcccccm{130pt}}
        \hline
        \textbf{Universities} & \textbf{oai\_dc\ } & \textbf{qdc\ } & \textbf{oai\_qdc\ } & \textbf{etdms\ } & \textbf{oai\_etdms\ } & \textbf{others}\\
        \hline
        Acadia University & X &   & X &   & X & mods\\
        Athabasca University & X &   &   &   & X & \\
        Bishop’s University* & X &   &   &   &   & didl, mets, oai\_bibl, rdf, uketd\_dc\\
        British Columbia Institute of Technology & X &   &   &   & X & mods\\
        Brock University & X & X &   & X &   & didl, dim, marc, mets, mods, ore, rdf, uketd\_dc, xoai\\
        Cape Breton University & X &   & X &   & X & mods\\
        Capilano University & X &   & X &   & X & mods\\
        Carleton University & X &   &   &   &   & mods\\
        Collège Militaire Royal du Canada & X & X &   & X &   & didl, dim, marc, mets, mods, ore, rdf, uketd\_dc, xoai\\
        Dalhousie University & X & X &   & X & X & didl, dim, marc, mets, mods, ore, rdf, uketd\_dc, xoai\\
        Emily Carr University of Art and Design & X &   & X &   & X & mods\\
        HEC Montréal & X &   &   &   &   & inmedia\\
        Institut National de la Recherche Scientifique & X &   &   &   &   & didl, mets, oai\_bibl, rdf, uketd\_dc\\
        Kwantlen Polytechnic University & X &   & X &   & X & mods\\
        Lakehead University & X & X &   & X &   & didl, dim, marc, mets, mods, ore, rdf, uketd\_dc, xoai\\
        MacEwan University & X & X &   & X &   & didl, dim, marc, mets, mods, ore, rdf, uketd\_dc, xoai\\
        McMaster University & X & X &   & X &   & didl, dim, lac2, lac\_mac, marc, mets, mods, ore, rdf, uketd\_dc, xoai\\
        Memorial University of Newfoundland & X &   &   &   &   & didl, etd-ms, mets, oai\_bibl, rdf, uketd\_dc\\
        Mount Royal University & X &   & X &   & X & mods\\
        Mount Saint Vincent University & X & X &   & X &   & didl, dim, marc, mets, mods, ore, rdf, uketd\_dc, xoai\\
        NSCAD University & X &   & X &   & X & mods\\
        National School of Public Administration & X &   &   &   &   & didl, mets, oai\_bibl, rdf, uketd\_dc\\
        OCAD University & X &   &   &   & X & didl, mets, oai\_bibl, rdf, uketd\_dc\\
        Ontario Tech University & X & X &   & X & X & didl, dim, marc, mets, mods, ore, rdf, uketd\_dc, xoai\\
        Queen's University & X & X &   & X &   & didl, dim, marc, mets, mods, ore, rdf, uketd\_dc, xoai\\
        Royal Roads University & X & X &   & X &   & didl, dim, marc, mets, mods, ore, rdf, uketd\_dc, xoai\\
        Saint Mary's University & X & X &   & X &   & didl, dim, marc, mets, mods, ore, rdf, uketd\_dc, xoai\\
        Simon Fraser University* & \\
        St. Francis Xavier University & X &   & X &   & X & mods\\
        TELUQ University & X &   &   &   & X & didl, mets, oai\_bibl, rdf, uketd\_dc\\
        Thompsom Rivers University & X &   & X &   & X & mods\\
        Trinity Western University & X &   & X &   & X & mods\\
        University of Alberta & X &   &   &   & X & \\
        Université Concordia d’Edmonton & X &   &   &   & X & didl, mets, oai\_bibl, oai\_openaire, oai\_ore\_atom, oai\_ore\_rdf, rdf, uketd\_dc\\
        University of Calgary & X & X &   & X &   & didl, dim, marc, mets, mods, ore, rdf, uketd\_dc, xoai\\
        University of Guelph & X & X &   & X & X & didl, dim, marc, mets, mods, ore, rdf, uketd\_dc, xoai\\
        University of Manitoba & X & X &   & X & X & didl, dim, marc, mets, mods, oai\_lockss, ore, rdf, uketd\_dc, xoai\\
        University of Northern British Columbia & X &   & X &   & X & mods\\
        University of Prince Edward Island & X &   & X &   & X & mods\\
        University of Quebec, Abitibi-Temiscamingue & X &   &   &   &   & didl, mets, oai\_bibl, rdf, uketd\_dc\\
        University of Quebec at Chicoutimi & X &   &   &   &   & didl, oai\_bibl, uketd\_dc\\
        University of Quebec in Montreal & X &   &   &   &   & didl, mets, oai\_bibl, rdf, uketd\_dc\\
        University of Quebec,  Trois-Rivieres & X &   &   &   & X & didl, etd\_ms\_uqtr, mets, oai\_bibl, oai\_openaire, rdf, uketd\_dc\\
        University of Regina & X & X &   & X &   & didl, dim, marc, mets, mods, ore, rdf, uketd\_dc, xoai\\
        University of Saskatchewan & X & X &   & X &   & didl, marc, mets, mods, ore, rdf, uketd\_dc\\
        University of Toronto & X & X &   & X & X & didl, dim, marc, mets, mods, ore, rdf, uketd\_dc, xoai\\
        University of Victoria & X & X &   & X & X & didl, dim, marc, mets, mods, oai\_etdms\_old, oai\_openaire, ore, rdf, uketd\_dc, xoai\\
        University of Waterloo & X & X &   & X &   & didl, dim, marc, mets, mods, ore, rdf, uketd\_dc, xoai\\
        University of Winnipeg & X & X &   & X & X & didl, dim, marc, mets, mods, ore, rdf, uketd\_dc, xoai\\
        University of the Fraser Valley & X &   & X &   & X & mods\\
        Université Laurentienne & X & X &   & X & X & didl, dim, marc, mets, mods, ore, rdf, uketd\_dc, xoai\\
        Université Laval & X & X &   & X &   & didl, dim, etdms11, marc, mets, mods, ore, rdf, uketd\_dc, xoai\\
        Université McGill* & &&&&& \\
        Université de Lethbridge* & X & X &   & X & X & didl, dim, marc, mets, mods, ore, rdf, uketd\_dc, xoai\\
        Université de Moncton & X &   & X &   & X & mods\\
        Université de Montréal & X & X &   & X & X & didl, dim, etdms10, etdms11, marc, marc21, mets, mods, oai\_openaire, oai\_openaire4science, ore, rdf, uketd\_dc, xoai\\
        Université de Sherbrooke & X & X &   &   & X & didl, dim, marc, mets, mods, ore, rdf, uketd\_dc, xoai\\
        Université d’Ottawa & X & X &   & X & X & didl, dim, marc, mets, mods, ore, rdf, uketd\_dc, xoai\\
        York University & X & X &   & X & X & didl, dim, marc, mets, mods, musicmods, ore, rdf, uketd\_dc, xoai\\
        Western University* & X & X &   &   & X & dcq, dcs, oai-dc, qualified-dublin-core, simple-dublin-core\\
        École Polytechnique de Montréal & X &   &   &   & X & didl, mets, oai\_bibl, oai\_openaire, rdf, rem\_atom\\
        École de Technologie Supérieure & X &   &   &   &   & didl, mets, oai\_bibl, rdf, uketd\_dc\\
        \hline
    \end{tabular}}
    \label{tab:SupportedFormats}
\end{table} \vspace{-5pt}

\newpage
\section{Date ISO distribution and formats}
\label{app:date}

\begin{figure}[H]
    \centering
    \includegraphics[width=1.0\textwidth]{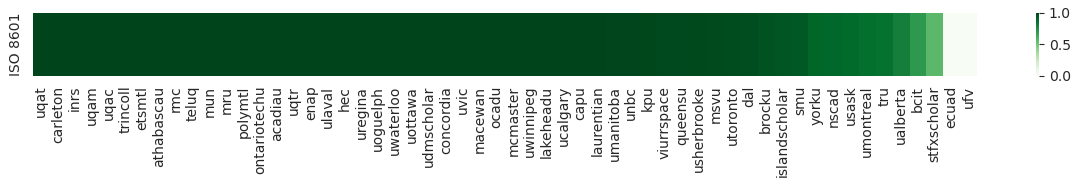}
    \caption{Proportion of dates in an ISO 8601 format in each university repository (we discard any repository that never includes the date tag).}
    \label{fig:date_iso}
\end{figure}

\begin{figure}[H]
    \centering    \includegraphics[width=1.0\textwidth]{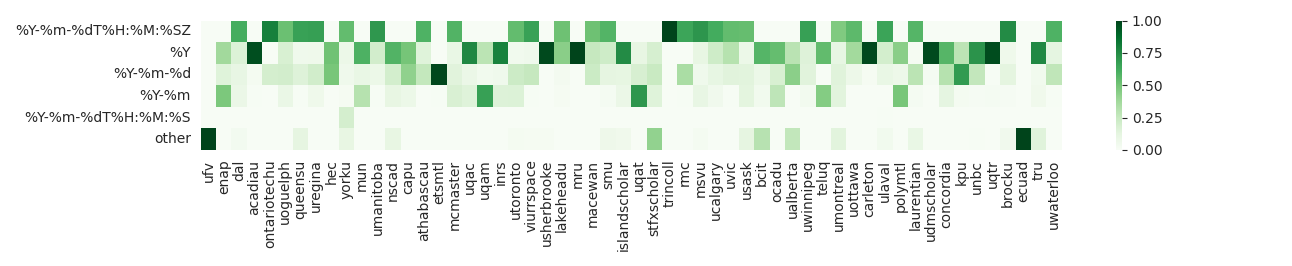}
    \caption{Distribution of most used date formats in each university repository (we discard any repository that never includes the date tag).}
    \label{fig:date_form}
\end{figure}

\section{Type Regular expressions}
\label{app:regexes}
\begin{center}
\begin{tabular}{lr}
    \textbf{Type} & \textbf{Regular expressions} \\
    \hline
    Article & \texttt{/*article*/i} \\
    Book & \texttt{/*(book)|(livre)*/i} \\
    Dataset & \texttt{/*dataset*/i} \\
    Media & \texttt{/*(photo)|(m[eé]dia)|(image)|(picture)|(vid[ée]o)|(audio)*/i} \\
    Poster & \texttt{/*poster*/i} \\
    Presentation & \texttt{/*pr[eé]sentation*/i} \\
    Thesis & \texttt{/*(thesis)|(th[eè]se)|(m[eé]moire)|(dissertation)*/i} \\
    \hline
\end{tabular}
\end{center}

\section{Language capturing Regular expressions}
\begin{table}[H]
    \caption{Regular expressions used to capture different variations of language names appearing in the \xmltag{Language} tag and map them to a unambiguous standard name. The Regular expressions are simple since the variety of languages mentioned in Canadian UIRs is quite limited and does not require to cover complex cases. All strings were lowercased previous to using the regular expressions. Our capturing script was written in Python and therefore, we used its native Traditional DFA (deterministic finite automaton) regular expression engine.}
    \centering
    \begin{tabular}{lrr}
        \textbf{Targeted language} & \textbf{Regular expressions} & \textbf{Clarification} \\
        \hline
        English & \texttt{/*(\^{}en)|(\^{}angl)*/i} & Matches a string whose first characters are \emph{en} or \emph{angl}.\\
        French & \texttt{/*(\^{}fr)*/i}  & Matches a string whose first characters are \emph{en}. \\
        \hline
    \end{tabular}
    \label{app:lang_regexes}
\end{table}








\end{document}